\documentclass[twocolumn,english]{revtex4-1}
\usepackage[T1]{fontenc}
\usepackage[latin9]{inputenc}
\usepackage{geometry}
\usepackage{float}
\usepackage{graphicx}
\usepackage{xcolor}
\usepackage{amsmath}

\usepackage{hyperref}
\hypersetup{
    colorlinks=true,    % akme links colored
    linkcolor=blue,     % internal links (like figures) will be blue
    urlcolor=blue,      % external urls will be blue
    citecolor=blue      % citations will be blue
}

\usepackage{babel}

\begin{document}

\title{Banding and polarization in driven multistable materials}

\author{Sheng Huang}
\email{sheng.huang@campus.technion.ac.il}
\affiliation {Faculty of Mechanical Engineering, Technion, Haifa 32000, Israel}

\author{Daniel Hexner}
\email{danielhe@technion.ac.il}
\affiliation {Faculty of Mechanical Engineering, Technion, Haifa 32000, Israel}

%%%%%%%%%%%%%%%%%%% ABSTRACT %%%%%%%%%%%%%%%%%%%
\begin{abstract}

We study a disordered network of bistable bonds subjected to periodic strain. The model is inspired by experiments on crumpled sheets and it features behaviors associated with glasses, including a complex energy landscape, memories, and large avalanches. At small strain amplitudes, the system quickly converges to a limit cycle where the system repeatedly cycles between a set of states. At large amplitudes, motion is erratic and does not converge to a limit cycle. The transition appears to be continuous, with diverging time scales. The nature of instabilities is different on both sides of the transition. At small strain amplitudes, instabilities are correlated only over a finite distance. Above the transition, instabilities are localized along diagonal bands. The distance between bands grows near the transition and appears to diverge. We propose a simple model that explains these observations. Below the transition, we propose a new ``order parameter'' -- the polarization of the instabilities along the driving direction. 

\end{abstract}
\maketitle

%%%%%%%%%%%%%%%%%%% INTRODUCTION %%%%%%%%%%%%%%%%%%%
\section{Introduction}

Subjecting materials to periodic drive often changes the microstructure~\cite{pine2005chaos, fiocco2013oscillatory, regev2013onset, mohan2013local,ong2024jamming,bhaumik2025yielding} and, as a result, also their mechanical properties~\cite{leishangthem2017yielding, keim2019memory, berthier2025yielding}. Crystalline solids undergo strain hardening~\cite{nix1971physics}, glasses may experience rejuvenation~\cite{lacks2004energy} or aging~\cite{hodge1995physical}, and suspensions change viscosity~\cite{franceschini2014dynamics}. Understanding the effect of driving could allow tuning material properties in a beneficial manner. 

Identifying how the microscopic state evolves due to driving is a central challenge. In amorphous materials, this is particularly challenging since there is no obvious apparent ordering. Nonetheless, experiments~\cite{shohat2022memory,ghosh2022coupled} and simulations~\cite{regev2013onset,leishangthem2017yielding} have demonstrated that periodic driving often entails large relaxation times, suggesting an underlying structural organization. 

We are inspired by studies on periodically driven non-Brownian suspensions~\cite{pine2005chaos,corte2008random}. There, a transition occurs as a function of the driving amplitude from an absorbing phase, where the motion of all particles is periodic, to a chaotic state at large amplitudes. Ordering occurs both on short length scales, where particles organize to avoid collisions~\cite{corte2008random}, as well as on long length scales, suppressing large-scale fluctuations~\cite{hexner2015hyperuniformity,tjhung2015hyperuniform}. Though disordered materials, such as glasses, undergo a similar transition~\cite{fiocco2013oscillatory,regev2013onset,bhaumik2022avalanches,ghosh2022coupled}, no clear ordering has yet been observed.

In this paper, we study a disordered bistable network~\cite{shohat2022memory, shohat2023logarithmic, shohat2025emergent, kedia2023drive} under periodic drive. The model is motivated by experiments on crumpled sheets, and has been successful in capturing various effects, including the occurrence of limit cycles~\cite{shohat2022memory}, memories~\cite{shohat2022memory}, avalanches~\cite{shohat2025emergent}, and aging~\cite{shohat2023logarithmic,shohat2025emergent}. This model features behaviors associated with glasses, but is more transparent since all instabilities can be enumerated. 

Through numerical simulations in the quasistatic regime, we study the transition as a function of the strain amplitude. At small strain amplitudes, the system converges to a limit cycle that periodically revisits the same states. At large amplitudes, the dynamics continually evolve. A sharp transition separates these two phases, with continuous characteristics, including diverging time scales, continuously varying ``activity'', and growing length scales.

The emergence of large time scales near the transition often suggests the presence of diverging length scales. We study the structural organization of the instabilities. Above the transition, instabilities form diagonal, system-spanning bands. The distance between bands grows when approaching the transition and appears to diverge. Below the transition, instabilities appear to be uncorrelated on long length scales.
We propose a simple variational model to explain the transition and the formation of bands. 

Lastly, we show another form of spatial organization. Instabilities become polarized along the direction of driving. We define several order parameters and show that they gradually grow, with time scales comparable to the relaxation of the activity. This suggests that it is an important ingredient for reaching a limit cycle.  Furthermore, this mechanism allows the energy of the deformation to decrease with driving and serves as a means of encoding a memory of the deformation.  

%%%%%%%%%%%%%%%%%%% MODEL %%%%%%%%%%%%%%%%%%%
\section{Model}

\begin{figure}[h]
    \begin{center}
    \includegraphics[width=1\columnwidth]{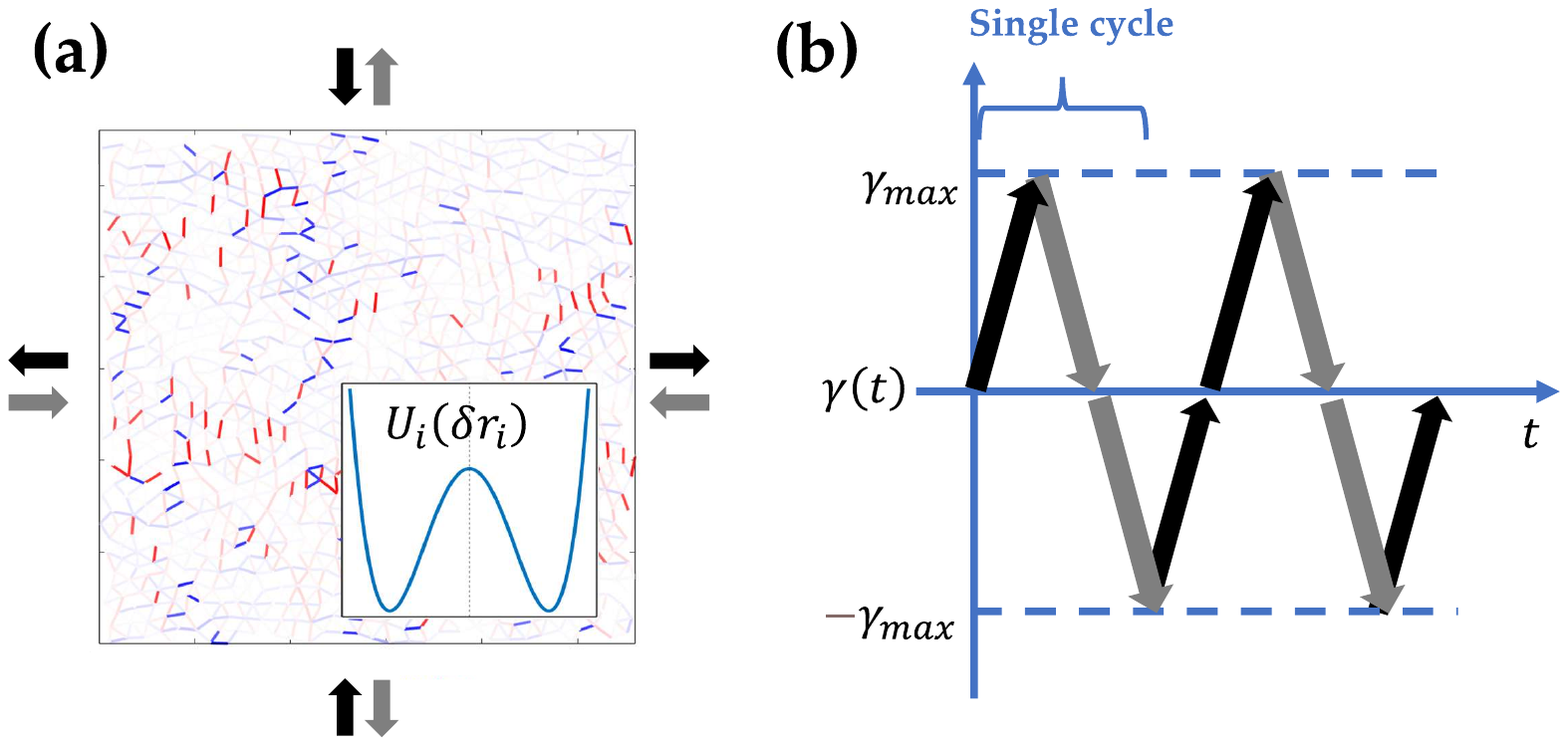}    
    \end{center}
    \caption{\textbf{An illustration of the model and shear protocol.} (\textbf{a}) A bistable network under pure shear. The inset shows the potential energy of a bistable bond [Eq. \eqref{eq:U}] as a function of displacement from $r_0$, the midpoint between the two minima. The color shading of the bond illustrates the extension between the maximal and minimal strain. Red denotes compression, while blue denotes elongation. (\textbf{b}) Periodic shearing strain protocol as a function of time. A single cycle consists of the strain sequence: $0 \to \gamma_{max} \to 0 \to -\gamma_{max} \to 0$.}
    \label{fig:Model}
\end{figure}

We study a disordered network of bistable springs~\cite{benichou2013structures,bertoldi2017harnessing,shohat2022memory,shuminov20242,sirote2024emergent,liu2024controlled} in two dimensions that is strained periodically. The networks are prepared from packings of soft spheres at zero temperature~\cite{Ohern2003jamming}. This ensemble is convenient since it allows the creation of networks with known properties, such as coordination number and local geometry. An example is shown in Figure~\hyperref[fig:Model]{{\ref{fig:Model}}(a)}. To avoid the influence of the jamming transition, we select the coordination number (twice the number of bonds per node) to be large, $Z \approx 5.5$, which is far from the jamming point $Z_c =2d=4$. The boundary conditions are taken to be periodic to avoid edge effects. 
    
To prepare the networks, we begin with a packing at force balance under an imposed pressure. The nodes are associated with the centers of the spheres, and overlapping particles are attached with a bond. The energy of each bond has two minima, with a short and a long state. For simplicity, we take the potential to be symmetric around $r_0$, which also defines the local energy maximum. The potential as a function of the displacement $\delta r$ from $r_0$ is given by,

\begin{equation}
    U_i = \frac{C}{4} \bigg (\frac{\delta r_i}{r_{0,i}} \bigg )^4 - \frac{\alpha}{2} \bigg (\frac{\delta r_i}{ r_{0,i}} \bigg )^2
    \label{eq:U}
\end{equation}
We choose for all bonds the same parameters, $C=1$, $\alpha=0.01$, and $r_0$ is taken from the initial distance between the nodes.  

In this study, we periodically shear the network up to a maximum shear strain $\gamma_{max}$, as illustrated in Figure~\hyperref[fig:Model]{{\ref{fig:Model}}(b)}. We apply a pure shear, where the x-dimension and y-dimension of the box are varied with an opposite phase. 
    
We focus on quasistatic actuation, where the system is assumed to be at force balance at any given time. To this end, we vary the strain iteratively and minimize the energy to reach force balance, using the FIRE algorithm~\cite{bitzek2006structural}. We take the number of iterations per cycle to be modest (typically 40 or 80). We show in Appendix \ref{Appendix: steps per cycle} that this has a negligible effect. Overall, the strains we consider are small and the strain steps do not exceed $\Delta \gamma \leq 0.002$.

%%%%%%%%%%%%%%%%%%% RESULTS %%%%%%%%%%%%%%%%%%%
\section{Results}

\subsection{A transition from limit cycles to an active phase}

We begin by demonstrating that this system undergoes a phase transition from having limit cycles at small strain amplitudes to an ``active'' phase where the system continually evolves at large strain amplitudes. We employ the nomenclature of absorbing phase transitions and define an ``activity'' that characterizes the change to the structure of the system from cycle to cycle. The activity is defined in terms of the change in bond length, $l_i$,
\begin{equation}
    a(\tau) = \frac{1}{N_b} \sum_{i=1}^{N_b} \bigg(l_i (\tau) - l_i (\tau-1)\bigg)^2 .
    \label{eq:a}
\end{equation}

\begin{figure}[h]
    \begin{center}
\includegraphics[width=1\columnwidth]{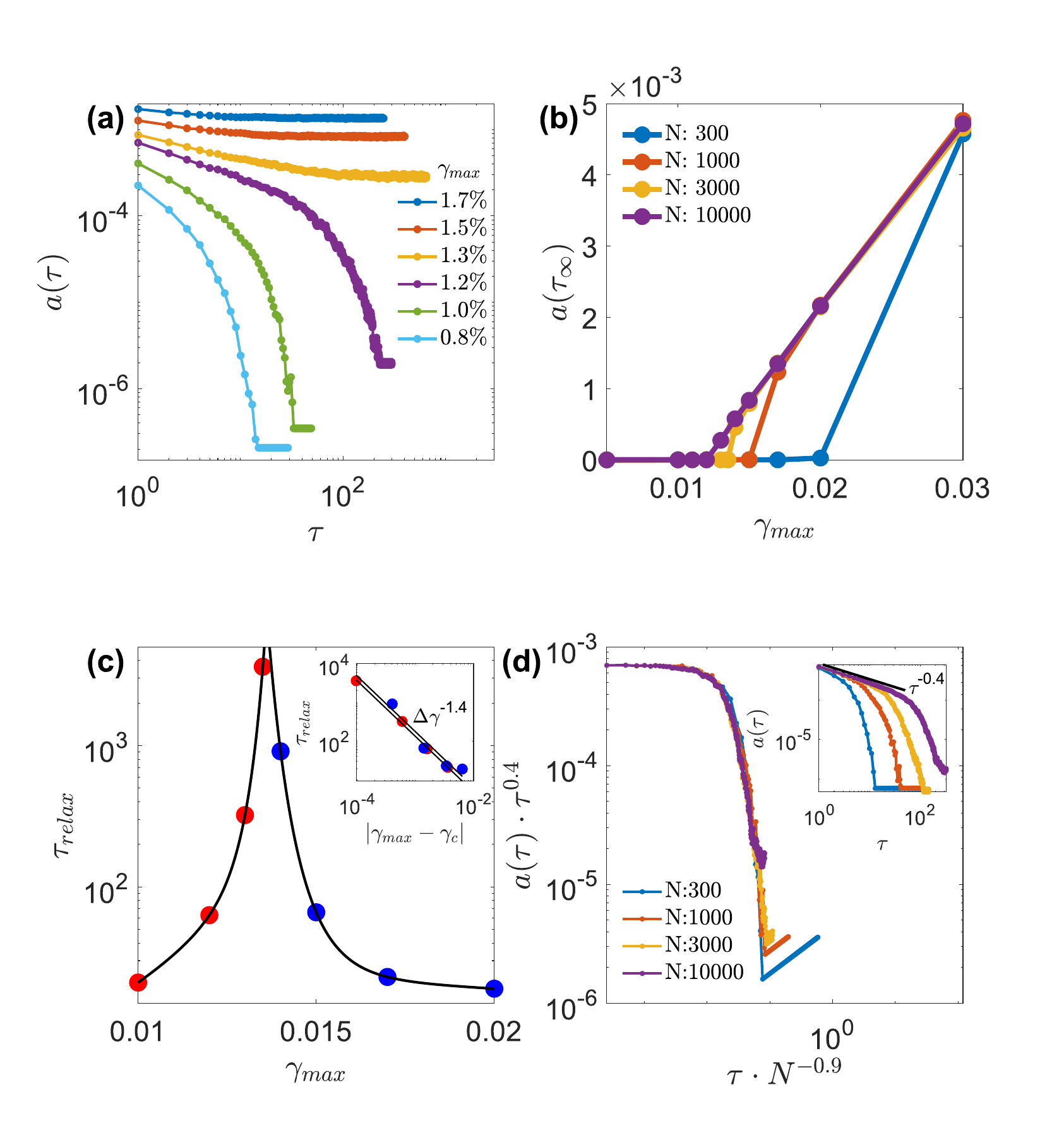}
    \end{center}
     \caption{\textbf{Characterizing the phase transition via stroboscopic activity.} (\textbf{a}) The stroboscopic activity $a(\tau)$ [Eq. \eqref{eq:a}] as a function of strain cycle $\tau$ for various driving amplitudes. For small amplitudes, $a(\tau)$ decreases gradually to zero. A small fraction of realizations converge to limit cycles where the period is multiple of the driving period. At large amplitudes, $a(\tau)$ converges to a nonzero value, indicating an active phase. (\textbf{b}) Activity after a large number of cycles $a(\tau_{\infty})$ as a function of $\gamma_{max}$, showing a sharp transition at $\gamma_c \approx 0.0127$. Finite-size effects are evident, with smaller systems exhibiting limit cycles even for $\gamma_{max} > \gamma_c$. (\textbf{c}) Relaxation time $\tau_{relax}$ as a function of $\gamma_{max}$. For $\gamma_{max} > \gamma_c$, $\tau_{relax}$ diverges near the transition. Inset: Log-log plot of $\tau_{relax}$ versus $|\gamma_{max} - \gamma_c|$, yielding a divergence exponent of approximately $-1.4$. (\textbf{d}) Finite-size scaling collapse of $a(\tau)$ near the critical strain amplitude $\gamma_c$.
    %revealing a system-size exponent $z \approx 0.9$ in $\tau \propto N^z$. Inset: Temporal decay exponent $\delta \approx 0.4$ in $a \propto \tau^{-\delta}$.
    }
    \label{fig:phase-transition-1}
\end{figure}

Figure~\hyperref[fig:phase-transition-1]{{\ref{fig:phase-transition-1}}(a)} shows the evolution of the activity $a(\tau)$ for various amplitudes, $\gamma_{max}$. For small amplitudes, the activity decreases and converges to a small value. Most realizations reach a limit cycle with a period equal to that of driving. A small fraction of the realizations converge to limit cycles with a large period. At larger amplitudes, the activity converges to a non-zero value.

Figure~\hyperref[fig:phase-transition-1]{{\ref{fig:phase-transition-1}}(b)} shows the activity after a large number of cycles. At small amplitudes, the activity is small (zero if we discard the limit cycles with periods larger than one), while at large amplitudes the activity appears to grow approximately linearly, $a \left( \tau_{\infty} \right)\propto \left| \gamma_{max} - \gamma_c \right|^{\approx 1.0}$. We note that there are significant finite-size effects. For small systems, the system reaches a limit cycle even for $\gamma_{max}>\gamma_c$. However, as the system size increases, the threshold for having active evolution converges to $\gamma_c \approx 0.0127$. That is, there appears to be a well-defined sharp transition in the thermodynamic limit with a critical strain that is system size independent. Similar finite-size effects are common in absorbing transitions~\cite{henkel2008non}.

We also demonstrate in Figure~\hyperref[fig:phase-transition-1]{{\ref{fig:phase-transition-1}}(c)} that there is a diverging time scale, associated with the number of cycles needed to reach a limit cycle at small $\gamma_{max}$ and the time to reach a steady state at large $\gamma_{max}$. For $\gamma_{max} > \gamma_c$ we estimate the relaxation time by measuring the number of cycles needed for the activity to decrease to $1/1000$ of the steady state value. Other thresholds give similar results. 

Figure~\hyperref[fig:phase-transition-1]{{\ref{fig:phase-transition-1}}(c)} shows that the time scale diverges as the strain amplitude approaches the critical value. In the inset we estimate that $\tau_{relax} \propto \left| \gamma_{max} - \gamma_c \right|^{\approx-1.4}$.  This exponent is similar to that found in directed percolation in two dimensions, $\nu_\parallel \approx 1.295$~\cite{grassberger1996self,lubeck2004universal}, but differs from the value reported for driven packings.~\cite{regev2013onset}. Lastly, in Figure~\hyperref[fig:phase-transition-1]{{\ref{fig:phase-transition-1}}(d)} we collapse the activity near the critical point for different system sizes and find that $\tau \propto N^{\approx 0.9}$. Again this is consistent with directed percolation~\cite{lubeck2004universal} where, $\tau\propto N^{\approx0.8830}$. At criticality, the activity decreases as $\tau^{\approx -0.4}$, which is similar to the directed percolation exponent $\alpha \approx0.4505$.

Altogether, the phenomenological behavior is consistent with a continuous-like absorbing transition. Some of the exponents are similar to those of directed percolation, while others are not. In particular, the exponent $\beta$ which characterizes the steady state activity. Below we discuss this in greater detail.
%, in particular, for $\gamma_{max}<\gamma_c$. We explore the changes to the structure with the number of periods and changing $\gamma_{max}$.

\begin{figure}[h]
    \begin{center}
    \includegraphics[width=1\columnwidth]{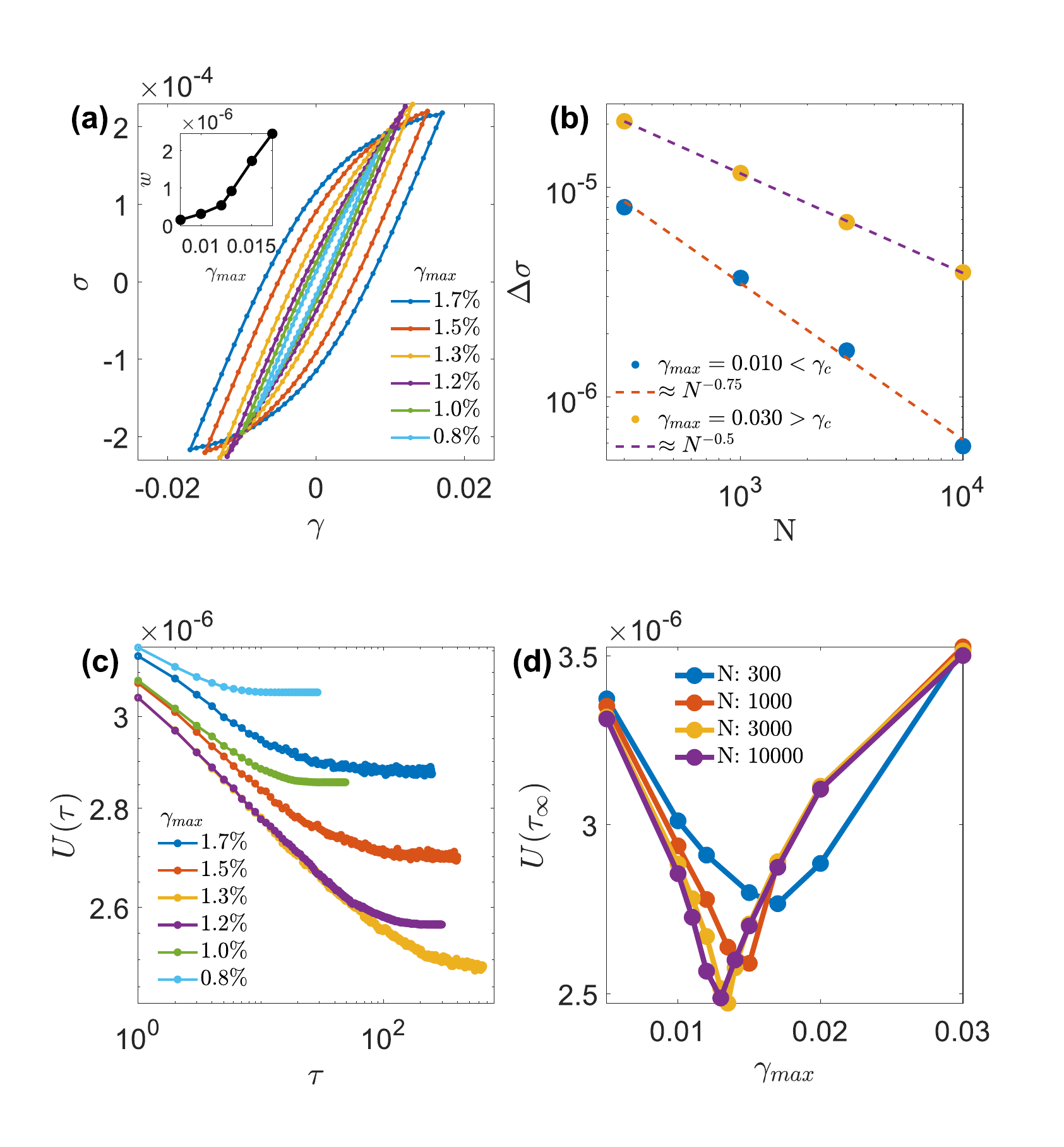}
    \end{center}
    \caption{\textbf{Mechanical signature of transition} (\textbf{a}) Stress $\sigma$ - strain $\gamma$ curves over a complete cycle for $\tau > \tau_\infty$, illustrating the hysteresis loop and its widening with the driving amplitude $\gamma_{max}$. Inset: The absorbed work $w$ during a cycle as a function of $\gamma_{max}$. (\textbf{b}) The average stress drop $\Delta \sigma$ as a function of system size for driving amplitudes above and below $\gamma_c$, showing distinct scaling behaviors corresponding to the limit cycle and active phases. (\textbf{c}) Temporal evolution of the potential energy at zero strain. The energy approximately decays logarithmically and then reaches a plateau. 
    %[Eq. \eqref{eq:U}] near the transition, highlighting two phases: convergence to limit cycles at small amplitudes and continued exploration of new states at large amplitudes. 
    (\textbf{d}) Potential energy after a large number of cycles $U(\tau_\infty)$ as a function of $\gamma_{max}$. The energy is minimal near the transition.}
    \label{fig:phase-transition-2}
\end{figure}

Next, we examine the effect of the transition on the elastic properties. Figure~\hyperref[fig:phase-transition-2]{{\ref{fig:phase-transition-2}}(a)} shows the stress-strain curves measured in a cycle after reaching a steady state. The stresses are computed using the virial stress equations~\cite{allen2017computer}. The stress-strain curves form hysteresis loops both below and above the transition. With increasing strain amplitude, the loops become broader, enclosing a larger area. The area corresponds to the work absorbed by the network during a cycle, $w$, resulting from the bistable transitions. Note that previous studies on driven glasses report that plasticity is negligible in the small-strain regime~\cite{leishangthem2017yielding}. As $\gamma_{max}$ grows, $w$ increases continuously throughout the transition, as shown in the inset of Figure~\hyperref[fig:phase-transition-2]{{\ref{fig:phase-transition-2}}(a)}. The transition is not easily discernible from the stress-strain curves; near the transition, there is a kink in the work as a function of $\gamma_{max}$. 
        
The stress-strain curves in Figure~\hyperref[fig:phase-transition-2]{{\ref{fig:phase-transition-2}}(a)} are smooth because they represent an average over many ensembles. However, in individual realizations, the stress-strain curves show discrete stress drops throughout a cycle. To quantify this, we calculate the average stress drop, $\Delta\sigma$. To this end, we strain with very small step sizes and compare the stress at each step to that of the previous one. A stress drop is identified when the stress changes with an opposite sign of the applied strain. The average stress drop per cycle is then computed once the system reaches a steady state.

Figure~\hyperref[fig:phase-transition-2]{{\ref{fig:phase-transition-2}}(b)} shows the average stress drop as a function of system size. We select two values of $\gamma_{max}$: one above $\gamma_c$ and one below $\gamma_c$. The results exhibit two distinct scaling behaviors, corresponding to two different phases. Above the transition,$\Delta \sigma \propto N^{\approx- 0.5}$, suggesting events whose dimension is $\sqrt{N}$ that scale with the system length. Recall that the contribution of a single bond to the virial stress scales inversely with volume. This scaling, we argue below, is associated with shear band-like events.  Below the transition, the scaling exponent, $\Delta \sigma \propto N^{\approx- 0.75}$ is somewhat surprising since it implies large correlated events. We interpret this exponent using results of Ref.~\cite{shohat2025emergent}, which showed that the system is marginal with avalanches that scale as $N^{\approx 0.27}$, yielding stress drops that scale as $\Delta \sigma \propto N^{\approx -0.73}$.

Next, we consider the evolution of the potential energy per particle. Figure~\hyperref[fig:phase-transition-2]{{\ref{fig:phase-transition-2}}(c)} illustrates the temporal evolution of $U(\tau)$ near the transition. Two distinct phases are observed: At small amplitudes, the $U(\tau)$ ceases to evolve after reaching a limit cycle. At large amplitudes, the $U(\tau)$ continues to evolve and fluctuate. The finite size analysis of energy evolution is shown in Appendix \ref{Appendix: finite size effects near the transition}. Interestingly, Figure~\hyperref[fig:phase-transition-2]{{\ref{fig:phase-transition-2}}(d)} reveals that $U(\tau_\infty)$, the energy after a large number of cycles, is minimal at the transition. This is similar to the behavior observed in periodically sheared glasses~\cite{leishangthem2017yielding}. The decrease in energy is an important phenomenon, which could be a governing principle for driven disordered materials~\cite{chvykov2021low}. Below, we also use this observation as a basis for theoretical arguments.

\subsection{Spatial organization for $\gamma>\gamma_c$}

\begin{figure}[h]
    \begin{center}
    \includegraphics[width=1\columnwidth]{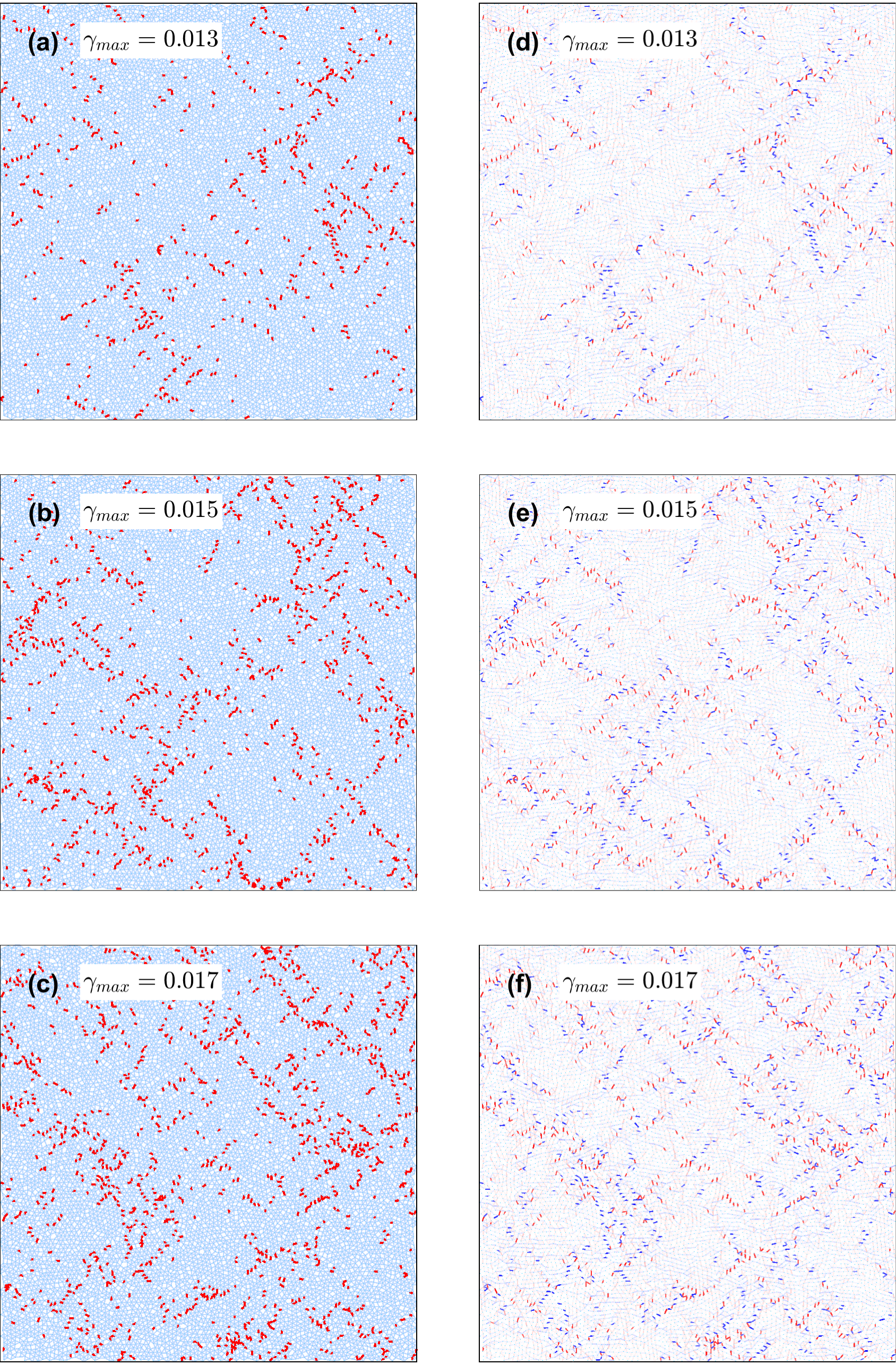}
    \end{center}
    \caption{\textbf{Diagonal bands spanning the system for $
    \gamma_{max}>\gamma_c$.} In (\textbf{a}), (\textbf{b}), and (\textbf{c}), the red bonds transition when the strain is varied from $-\gamma_{max}$ to $\gamma_{max}$, while the blue bonds did not. In (\textbf{d}), (\textbf{e}), and (\textbf{f}), the change in bond extension is shown. Red indicates compression while blue indicates elongation. The color intensity indicates the magnitude of the change.}
    \label{fig:visualization-above-transition}
\end{figure}

In continuous transitions, diverging relaxation times result from a diverging length scale. This leads us to ask if the large relaxation times, here, can be identified with a growing length scale in the structure? For reference, we compare to the directed percolation model, which is a prototypical model for absorbing transitions. There, the active phase has a diverging length scale associated with the distance between active regions (or the size of active regions). Unlike directed percolation, here there is an underlying structure, and we wish to understand if there is any signature in the structure.

Correlations in the structure of disordered materials are often not easily discernible~\cite{cavagna2009supercooled,hexner2015hyperuniformity,hexner2018two}. Here, we focus on the \emph{transitioning bonds}, defined as bonds that flip their state when the system is strained from $-\gamma_{max}$ to $\gamma_{max}$. That is, these are the bonds that undergo an instability or a rearrangement. Figure~\hyperref[fig:visualization-above-transition]{{\ref{fig:visualization-above-transition}}(a)}-\hyperref[fig:visualization-above-transition]{(c)} show snapshots of rearrangements at different strain amplitudes in steady state. The strain amplitudes are above the critical value in the active phase. In Figure~\hyperref[fig:visualization-above-transition]{{\ref{fig:visualization-above-transition}}(d)}-\hyperref[fig:visualization-above-transition]{{\ref{fig:visualization-above-transition}}(f)}, the elongation of the bonds is shown for the same strain amplitudes.  Blue denotes extension and red denotes compression.

Figure~\hyperref[fig:visualization-above-transition]{{\ref{fig:visualization-above-transition}}} indicates that the local strain and rearrangements are not uniformly distributed. Instead, they localize along bands $\pm 45^\circ$, similar to shear bands~\cite{maloney2006amorphous,manning2007strain,karmakar2010statistical,dasgupta2012microscopic,gendelman2013yield,alix2018shear,karimi2018correlation,rosner2024unveiling,berthier2025yielding}. These angles correspond to the principal directions where the shear stress is largest. As approaching the transition the distance between bands grows, and appears to diverge.

\begin{figure}[h]
    \begin{center}
    \includegraphics[width=1\columnwidth]{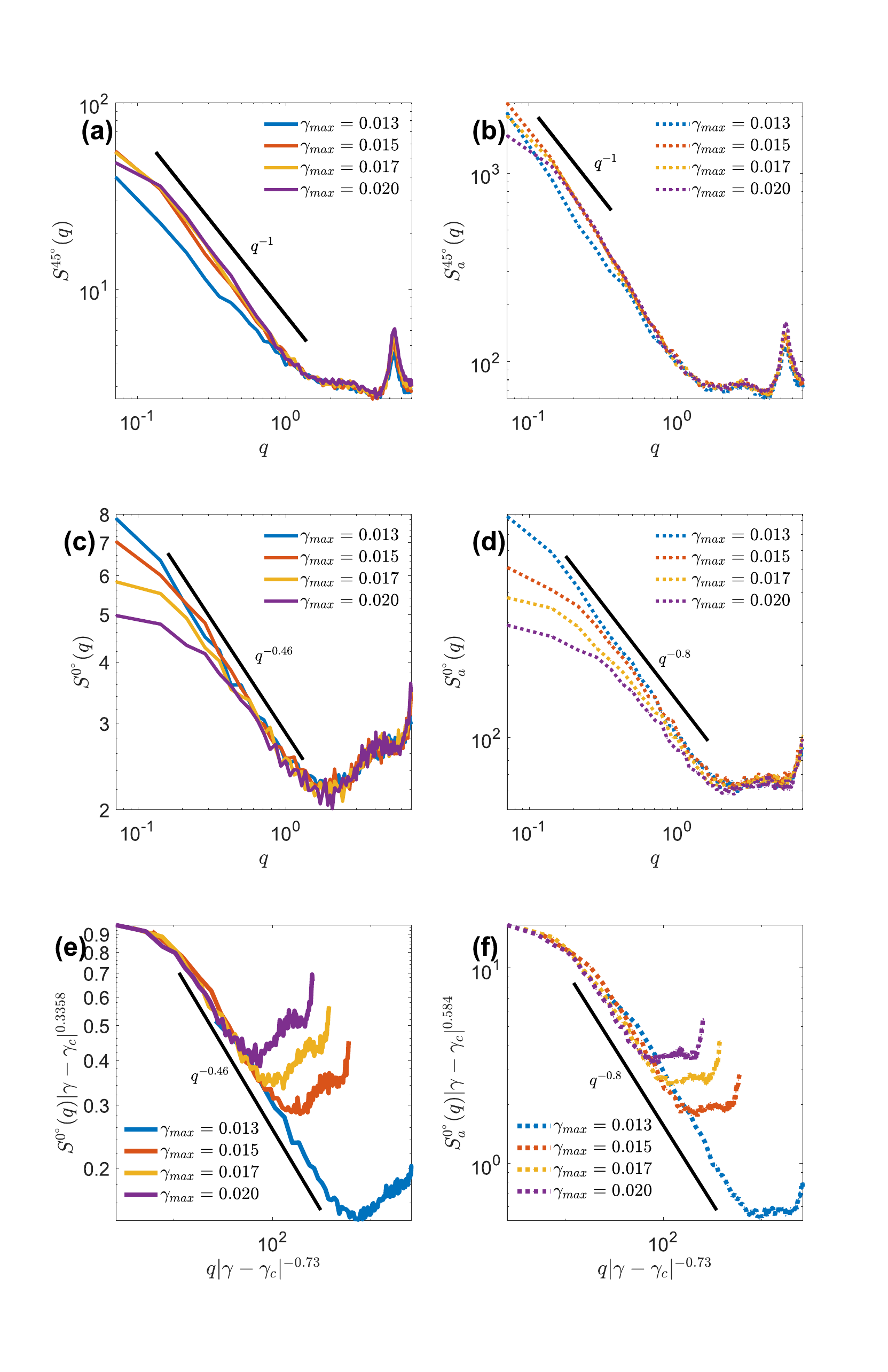}
    \end{center}
    \caption{\textbf{Structure factor in the active phase $\gamma_{max}>\gamma_c$}. Left: structure factor of the transitioning bonds. Right: structure factor of the activity. The power-law divergence of the structure factor in (\textbf{a}) and (\textbf{b}) along $45^\circ$ indicates that the bands persist over large distances. In (\textbf{c}) and (\textbf{d}), the structure factor along  $0^\circ$ measures the distance between bands. The crossover from power-law scaling to a constant at small q defines a length scale that appears to grow. In (\textbf{e}) and (\textbf{f}) we show a data collapse aimed at measuring the growth of the length scale as approaching the transition. }
    \label{fig:SQ-TC-above-transition}
\end{figure}

To quantify the bands and the distance that separates the bands, we wish to define an appropriate correlation function. Correlation functions are difficult to measure over long distances, since they require a large number of realizations to resolve long-distance features. Instead, we measure the structure factor,
\begin{equation}
    S(q) = \frac{1}{N} \bigg \langle \bigg|\sum_{i=1}^{N_b} b_i e^{-i q r_i} \bigg|^2 \bigg \rangle.
    \label{eq:Sq}
\end{equation}

Here, $b_i$ can be any quantity, which in our case we take to be either the bond elongations, the indicator function for transitioning bonds ($b_i=1$ for transitioning bonds and zero otherwise), or activity. The structure factor is related to the correlation function via a Fourier transform:
\begin{equation}
    % S(q) =  \left\langle b_i^2 \right\rangle + \rho\int d^dr e^{-iq\cdot r} \left\langle b(r) b(0) \right\rangle - \left\langle b(0) \right\rangle^2
    S(q) =  \frac{1}{N} \left\langle \sum_{i=1}^{N_b} b_i^2 \right\rangle + \rho \int d^dr  \left\langle b(r) b(0)  \right\rangle e^{-iq\cdot r}
    \label{eq:Sq-Cr}
\end{equation}
where $q$ is the wave vector, $r$ is the position and $\rho$ is the number density. Appendix \ref{Appendix: finite size effects near the transition} shows that finite-size effects are small.

Figure~\hyperref[fig:SQ-TC-above-transition]{{\ref{fig:SQ-TC-above-transition}}} shows the structure factor of the transitioning bonds (left) and of the activity (right), defined by the change in bond length over a cycle. Recall that the wave vector $q$ is a vector with a direction. There are two important directions we consider. Measuring along $45^\circ$ characterizes the persistence of the band, while along $0^\circ$, $90^\circ$ measures the distance between bands. 

Figure~\hyperref[fig:SQ-TC-above-transition]{{\ref{fig:SQ-TC-above-transition}}(a)} and \hyperref[fig:SQ-TC-above-transition]{(b)} shows the structure factor measured along $45^\circ$ for different strain amplitudes above the transition. The structure factor approximately diverges as $q^{-1}$. This implies that the bands persist over a long distance. These long-range correlations, together with the stress drops that scale as $N^{-0.5}$, are evidence that the bands are system-spanning. We note that similar long-range correlations have been measured in sheared glasses~\cite{maloney2009anisotropic}.
 
Next, we consider the structure factor measured along $0^\circ$, with the aim of measuring the distance between bands. Figure~\hyperref[fig:SQ-TC-above-transition]{{\ref{fig:SQ-TC-above-transition}}(c)} and \hyperref[fig:SQ-TC-above-transition]{(d)} shows that the structure factor scales as $q^{\approx-0.46}$ and $q^{\approx-0.8}$ and then tapers off to a constant. The scaling $q^{2-\eta}$ corresponds to power-law correlations $r^{-d+2-\eta}$,  and the q-independent regime at small q-values indicates the loss of correlations on long length scales. The crossover between these two regimes defines the inverse correlation length. Note that at the critical point of directed percolation, the structure factor of the activity scales as $q^{\approx-0.4088}$, which is comparable to the exponent for the transitioning bonds. 
        
The shift of the crossover $q$ to smaller values indicates a growing correlation function, which we believe diverges. To characterize the divergence of the correlation length, we collapse the curves by rescaling the x and y axes with powers of $\gamma-\gamma_c$. We assume that the structure factor has a scaling form at small values $q^{2-\eta}f(q|\gamma-\gamma_c|^{-\nu})$. Since $\eta$ is directly measured through a fit, the collapse requires varying only a single exponent. 

Figure~\hyperref[fig:SQ-TC-above-transition]{{\ref{fig:SQ-TC-above-transition}}(e)} and \hyperref[fig:SQ-TC-above-transition]{(f)} that the correlation length or the typical distance between bands
scales as $\xi \propto |\gamma - \gamma_c|^{\approx-0.73}$. Interestingly, this exponent is consistent with the exponent of directed percolation~\cite{lubeck2004universal, hinrichsen2000non}.

\subsection{Spatial organization for $\gamma<\gamma_c$}

\begin{figure}[h]
    \begin{center}
    \includegraphics[width=1\columnwidth]{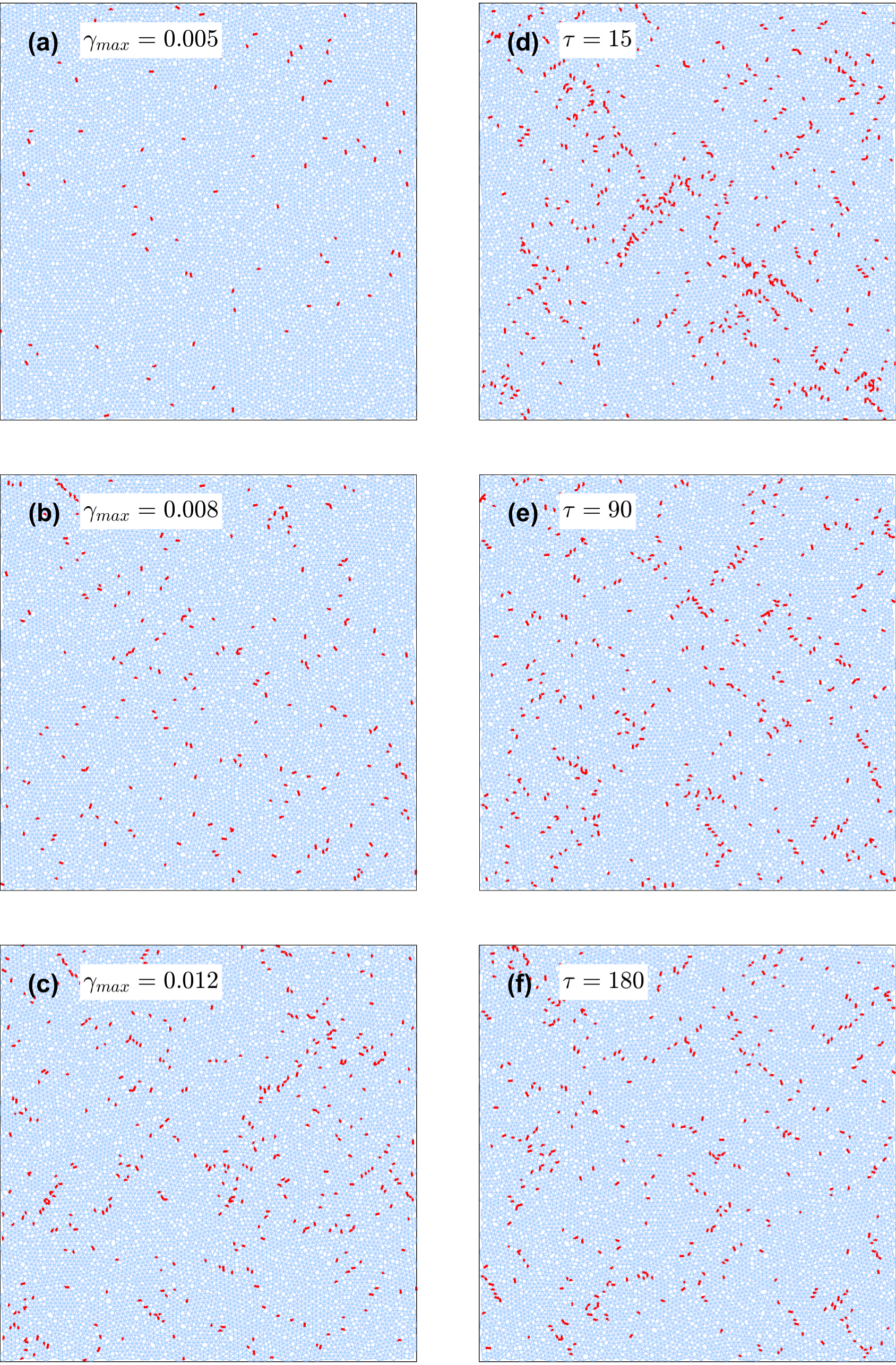}
    \end{center}
    \caption{\textbf{Structure of the transitioning bonds below the transition, $\gamma_{max}<\gamma_c$} Left: Example of the absorbing limit cycle phase at different strains. Right: Configuration at different times. In early times, there are bands that then disperse. }
    \label{fig:visualization-below-transition}
\end{figure}

We now examine the structural organization below the transition. Figure~\hyperref[fig:visualization-below-transition]{{\ref{fig:visualization-below-transition}}(a)-(c)} shows snapshots of systems that have reached limit cycles at strain amplitudes below the transition. When the strain amplitude is well below the transition, as in Figure~\hyperref[fig:visualization-below-transition]{{\ref{fig:visualization-below-transition}}(a)}, the rearrangements appear to be uniformly distributed. As the amplitude approaches the transition, as shown in Figure~\hyperref[fig:visualization-below-transition]{{\ref{fig:visualization-below-transition}}(b)} and \hyperref[fig:visualization-below-transition]{{\ref{fig:visualization-below-transition}}(c)}, some correlations are apparent.

\begin{figure}[h]
    \begin{center}
    \includegraphics[width=1\columnwidth]{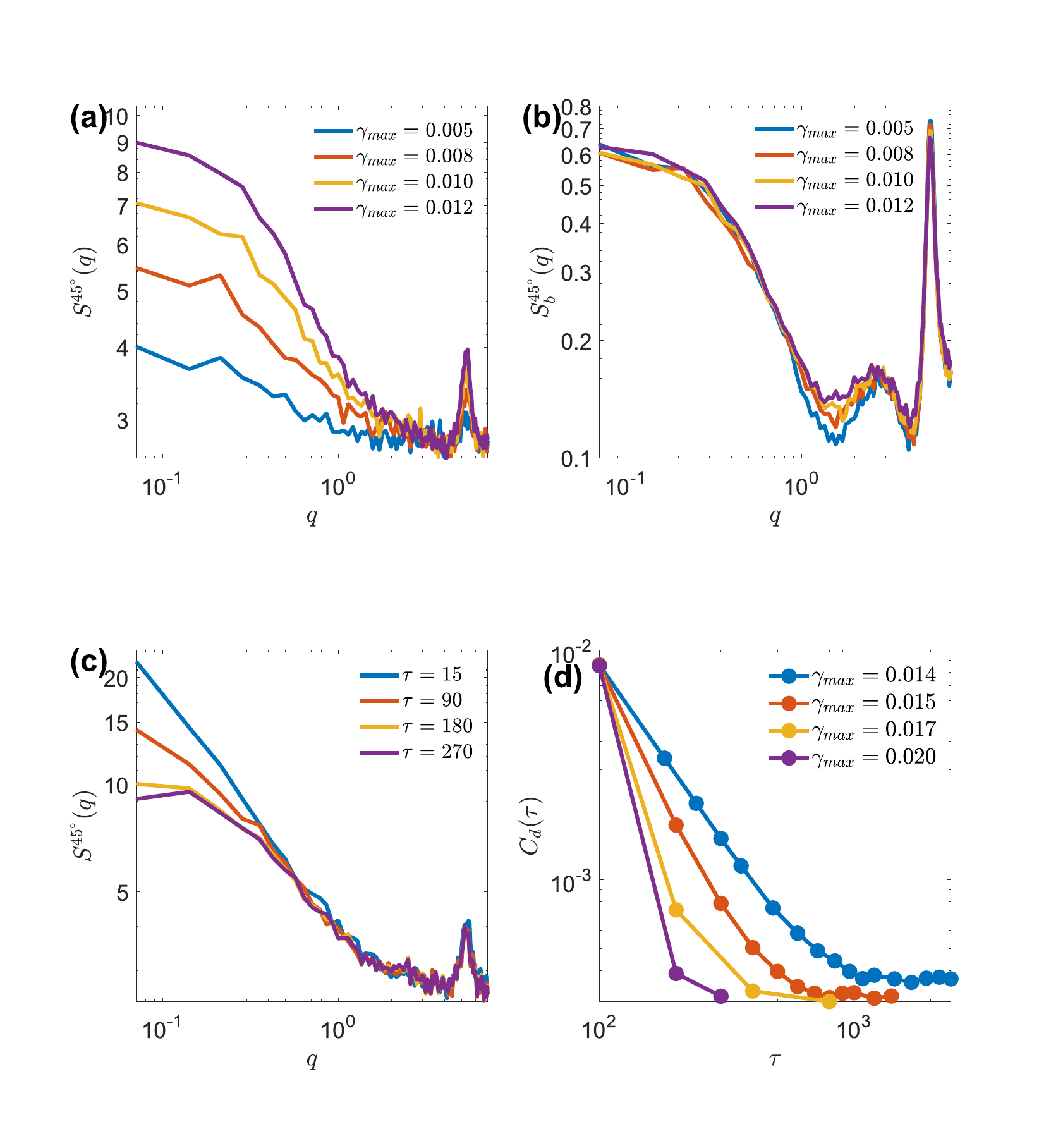}
    \end{center}
    \caption{\textbf{Structure factor below the transition $\gamma_{max}<\gamma_c$ (top) and temporal correlations (bottom).} Structure factor $S(q)$ measured along the $45^\circ$ direction of the transitioning bonds (\textbf{a}) and the bond elongations (\textbf{b})  for different strain amplitudes. There does not appear to be a diverging length scale.  (\textbf{c}) Structure factor measured along $45^\circ$ at different cycles during periodic drive, showing that the initial divergence associated with bands gradually disappears. This is consistent with the annealing of incipient bands in Figure \ref{fig:visualization-below-transition}. (\textbf{d}) Temporal correlation function $C_d(\tau)$ [Eq.~\eqref{eq:temp_corr}] for different $\gamma_{max}$. Relaxation slows down as the system approaches the transition, consistent with critical slowing down.}
    \label{fig:SQ-below-transition-TC}
\end{figure}

To quantify the correlations, we again measure the structure factor, focusing on the  $45^\circ$ direction. Figure~\hyperref[fig:SQ-below-transition-TC]{{\ref{fig:SQ-below-transition-TC}}(a)} shows the structure factor for the transitioning bonds, while Figure~\hyperref[fig:SQ-below-transition-TC]{{\ref{fig:SQ-below-transition-TC}}(b)} shows that structure factor for bond elongations. In both cases, the crossover length does not appear to grow; however, these two show different behavior. Figure~\hyperref[fig:SQ-below-transition-TC]{{\ref{fig:SQ-below-transition-TC}}(b)} shows that the structure factor for bond elongations exhibits nearly no dependence on $\gamma_{max}$. In contrast, the magnitude of the plateau of the structure factor, at small q values,  of the transitioning bonds grows with increasing $\gamma$. This growth appears to be continuous and non-divergent. The data indicates the correlation length does not diverge. The crossover length does not appear to vary much. 

We have experimented with varying different parameters, including the discretization and system size. We do not find any indication that there is a limit at which the structure factor diverges. There does seem to be a length scale in the problem $q\approx0.3$ corresponding to $\ell \sim 20$ (the typical length of a bond is order one). In the conclusions, we discuss this in greater detail.

\subsection{Temporal evolution}

We briefly discuss the temporal evolution, starting with the small-strain-amplitude phase. Figure~\hyperref[fig:visualization-below-transition]{{\ref{fig:visualization-below-transition}}(d)-(f)} shows the temporal evolution of the transitioning bonds just below the transition. Figure~\hyperref[fig:visualization-below-transition]{{\ref{fig:visualization-below-transition}}(d)} and \hyperref[fig:visualization-below-transition]{{\ref{fig:visualization-below-transition}}(e)}, show that at early times, diagonal bands form. However, as the number of applied cycles increases, the driving disrupts these bands, as shown in Figure~\hyperref[fig:visualization-below-transition]{{\ref{fig:visualization-below-transition}}(f)}. That is, periodic drive anneals out the bands. Figure ~\hyperref[fig:SQ-below-transition-TC]{{\ref{fig:SQ-below-transition-TC}}(c)} further supports this observation: the divergence of the structure factor measured along the \(45^\circ\), the direction of the bands, gradually vanishes with periodic drive.

Above the transition, we ask whether the bands are mobile. In Appendix
~\hyperref[Appendix: Supplementary_Video:evolution_of_bands]{{\ref{Appendix: Supplementary_Video:evolution_of_bands}}}
  we show a video of the transitioning bonds, which allows us to visualize the bands. Bands continually evolve and in a noisy, fluctuating manner. There is a clear distinction between regions that tend to accommodate bands and regions that are fairly frozen. Furthermore, there appears to be a second population of transitioning bonds that are \emph{distinct} from the bands. These transitioning bonds are mostly inactive and retain their state from cycle to cycle. Their state changes only after a large number of cycles.

To quantify the temporal evolution, we define the correlation function, 
\begin{equation}
    C_d(\tau)= \frac{1}{N} \sum_i \delta r_i (t+\tau)\delta r_i (t). 
    \label{eq:temp_corr}
\end{equation} 
Here, $t$ signifies a reference time when correlations begin to be measured. To study the steady state, $t$ is taken to be larger than the relaxation time. Figure~\hyperref[fig:SQ-below-transition-TC]{{\ref{fig:SQ-below-transition-TC}}(d)} shows the correlation function for different values of the strain amplitude. There is clear growth in relaxation time as approaching the transition. This is consistent with critical slowing down and signifies the slowdown of the motion of the bands.

\subsection{A variational model}

We propose a variational model to explain the transition. The response to an applied strain results in two contributions: (1) an elastic contribution, where the deformation on average is uniform throughout the system. (2) Instabilities that are localized along bands, which we refer to as the plastic contribution.  We assume that the system's energy is minimized due to the periodic drive. This is supported by observations that the energy decreases during training, especially near the transition (see  Figure~\hyperref[fig:phase-transition-2]{{\ref{fig:phase-transition-2}}(d)}).
        
Similar ideas on effective variational laws have been proposed for driven systems obeying damped Newton's equations~\cite{chvykov2021low,kedia2023drive}. There, it has been argued that the system evolves to minimize the work absorption. Work absorption injects energy into the system, allowing it to overcome barriers and explore the state space.  States with low work absorption are absorbing because they do not possess enough energy to escape from local minima. In quasistatics, transitioning between states depends on the internal energy (or stresses). States with a smaller internal energy (or stresses), we argue, explore the space of states to a lesser degree. 

The energy consists of an elastic contribution and a plastic contribution. The elastic energy, $U_{el}$, accounts for the approximately uniform deformations and is given by
\begin{equation}
    \frac{U_{el}}{L^2} = \frac{1}{2} G \gamma_{el}^2
\end{equation}
where $L$ is the system length, $G$ is the shear modulus and $\gamma_{el}$ is the elastic strain. The plastic energy, $U_{pl}$ of a band scales as the length of the system and the number of bands, $n_b$,
\begin{equation}
    U_{pl} = e_{pl} n_b L
\end{equation}
$e_{pl}$ is assumed to be the energy cost of a band per unit length. Each band contributes an extension of $\Delta$, and a strain (fractional change in length),
\begin{equation}
   \gamma_{pl} = \frac{n_b \Delta}{L} 
\end{equation}

As a result, the plastic energy density is linear in the plastic strain,
\begin{equation}
    \frac{U_{pl}}{L^2} =\frac{e_{pl}}{\Delta} \gamma_{pl}\end{equation}

The total energy density $u$ of the system is therefore,
\begin{equation}
    u = \frac{1}{2} G \gamma_{el}^2 + \frac{e_{pl}}{\Delta} \gamma_{pl}.
\end{equation}

Under our assumption, the state is determined by the minimum of $u$ with respect to $\gamma_{pl}$, with the constraint that the total strain $\gamma_{max} = \gamma_{el} + \gamma_{pl}$ is fixed. We discard the negative solution, since the plastic strain is positive by definition. This leads to the following prediction for the plastic strain:

\begin{equation}
\gamma_{pl} \propto \left\{
\begin{array}{ll}
|\gamma_{max} - \gamma_c|^1 & \gamma_{max} > \gamma_c \\
0 & \gamma_{max} < \gamma_c
\end{array} \right.
\end{equation}
Here, $\gamma_{c}=\frac{e_{pl}}{\Delta G}$. The distance between bands is given by
\begin{equation}
    \xi = \frac{L}{n_b} = \frac{\Delta}{\gamma_{pl}} \propto  |\gamma_{max} - \gamma_c|^{-1}
\end{equation}

The qualitative behavior is captured by the model, but it fails to capture the exponents quantitatively. This is expected since it is a mean-field approach. Moreover, the model neglects plasticity below the transition, effects of disorder, detailed spatial structure, and the marginal state of the system~\cite{shohat2025emergent}.

\subsection{Polarization}

\begin{figure}[h]
    \begin{center}
    \includegraphics[width=1\columnwidth]{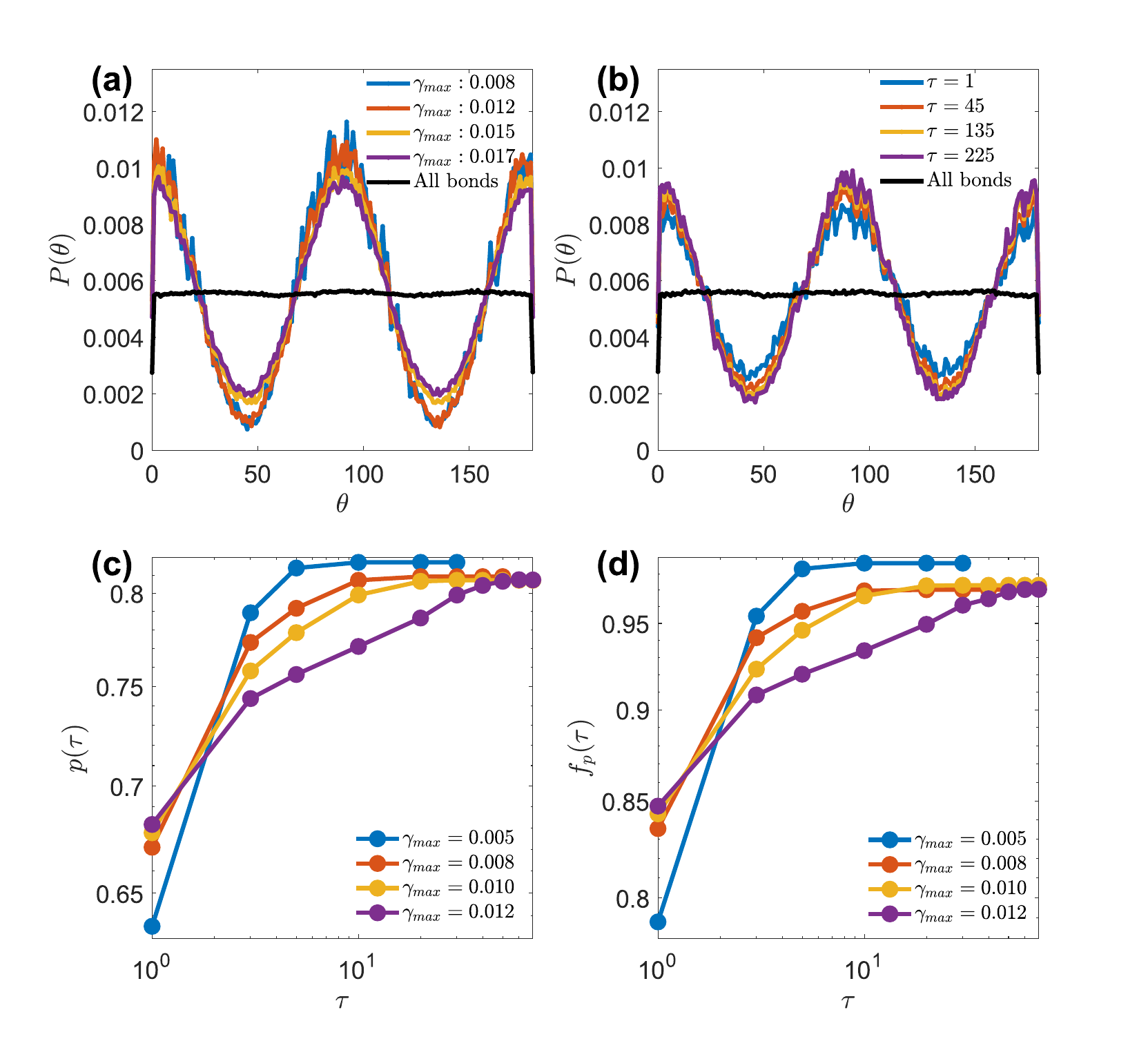}
    \end{center}
    \caption{\textbf{Polarization of transitioning bonds.} (\textbf{a}) Angular distribution of transitioning bonds reveals strong alignment along $0^\circ$ and $90^\circ$, in contrast to the uniform distribution observed when considering all bonds. (\textbf{b}) Degree of alignment increases with the number of cycles, indicating progressive polarization under periodic drive. (\textbf{c}) The polarization metric $p=\langle\cos(2\theta_i) \cdot \text{sign}(e_i)\rangle$ grows with time, suggesting alignment of bond elongation direction with the affine strain field. (\textbf{d}) An alternative measure, $f_p = \langle \text{sign}[\cos(2\theta_i)] \cdot \text{sign}[e_i] \rangle$, quantifies the fraction of bonds aligned with the affine deformation, reaching over 95\% at long times. Together, these results show that periodic driving induces structural polarization, which lowers the elastic energy and encodes memory of the applied deformation.}
    \label{fig:Polarization}
\end{figure}

Lastly, we turn to discuss another form of structural organization, which is also associated with the transitioning bonds.  The snapshots in Figure \ref{fig:visualization-above-transition} and Appendix \ref{Appendix: V PO} show that those bonds tend to align~\cite{lemaitre2021anomalous,livne2023geometric}  with the direction $0^\circ$ and $90^\circ$. To quantify this effect, we measure the distribution of the bond angle, shown in Figure~\hyperref[fig:Polarization]{{\ref{fig:Polarization}}(a)}. There is a clear alignment along the aforementioned angles.  For comparison, when including all the bonds, the distribution is uniformly distributed. Above the transition, there is slightly less alignment, but it is still significant. We rationalize this alignment by noting that the transiting bonds have a large contribution to the deformation. Therefore, they should align with the affine deformation.  Figure~\hyperref[fig:Polarization]{{\ref{fig:Polarization}}(b)} further demonstrates that this alignment progressively increases until reaching a limit cycle. This suggests that alignment plays a role in the formation of limit cycles.

Next, we wish to explore the relation between the change in length of the bond and its orientation. To include this information, we define different metrics for the polarization that encode the extension-angle relations. The affine extension of a bond is given by~\cite{lubensky2015phonons}, 
\begin{equation}
    e_{aff}=\sum_{i,j}\hat{r}_i\epsilon_{ij}r_j.
    \label{eq:affine_extension}
\end{equation}
Here, $r_i$ is the bond vector, and the strain tensor is that of pure shear. As a result, the affine extension is given by $e_{aff} =\epsilon r cos( 2\theta)$, where $r$ is the bond length and $\epsilon$ is the shear strain. 

We consider two measures of polarization, which highlight different aspects. In  Figure~\hyperref[fig:Polarization]{{\ref{fig:Polarization}}(c)} we plot $p=\frac{1}{N_b}\sum^{N_b}_{i=1}cos(2\theta_i) \cdot sign(e_i)$, which measures the correlation of the sign of elongation, $e_i$, with the orientation. This polarization measures the degree of angle polarization and is bound by unity, occurring if all bonds were oriented along $\theta=0^\circ,90^\circ$, and the elongation is perfectly correlated with the orientation. The polarization grows with time and reaches the near-maximal value $\approx 0.8$. The time scale on which this occurs grows as approaching the transition, indicating restructuring of the instabilities. 
        
We also define another metric for polarization that measures the fraction of the bonds that align with the affine deformation, $f_p=\frac{1}{N_b} \sum^{N_b}_{i=1}sign[cos(2\theta_i)] \cdot sign[e_i]$ (see Figure~\hyperref[fig:Polarization]{{\ref{fig:Polarization}}(d)}). The behavior is similar, and at large times over $95\%$ of the transitioning bonds are aligned with the deformation. 

In summary, periodic drive polarizes the transitioning bonds; a form of structural ordering. The strain polarizes the instabilities in analogy to the polarization of dipoles induced by a magnetic or electric field. Polarization reduces the elastic energy, since bonds that align with the affine deformation contribute to the deformation. Fewer transitioning bonds are required for the same value of strain. 

Polarization may be important in the formation of limit cycles. If bonds are all polarized, perhaps they interact in a coherent manner, or perhaps the strength of the interactions is reduced. 
Lastly, we note that polarization is a likely mechanism for encoding a memory of the applied external deformation

%%%%%%%%%%%%%%%%%%% Conclusions %%%%%%%%%%%%%%%%%%%

\section{Conclusion}

In conclusion, we have studied bistable networks under periodic drive. We have found that a transition occurs as a function of the strain amplitude between a phase traversing limit cycles to a phase that continually evolves and is characterized by banding. Bands occur at both $\pm45^\circ$. The distance between bands appears to diverge as it approaches the transition. The long-range correlations and the large stress drops are consistent with system-spanning bands.
    
Below the transition, the data indicates that there is no diverging length scale. A few comments are in order. Firstly, for comparison, in directed percolation, after reaching an absorbing state, there is no structural diverging length since the model has no underlying structure. Secondly, the model is known to be in a marginal state undergoing avalanches that diverge with system size~\cite{shohat2025emergent}. These we expect to manifest as diverging fluctuations, which, once again, we do not observe. There is some evidence that the system remains marginal even in the limit cycle phase~\cite{shohat2023logarithmic} (also see Figure~\hyperref[fig:phase-transition-2]{\ref{fig:phase-transition-2}(b)}), but further study is required.  

Interestingly, some of the exponents are similar to those of directed percolation. This includes the exponents $z,\nu_{\perp},\nu_{\parallel},\alpha$ corresponding to the finite size scaling, divergence of the relaxation time, and the divergence of the correlation length. The exponent $\eta$ that characterizes the power-law correlations of the structure factor is similar to the correlations of the transitioning bonds, but not that of the activity. Furthermore, the exponent $\beta\approx1$ is significantly different from that of directed percolation. 

We expect the transition to be different from that of directed percolation. Firstly, directed percolation is sensitive to disorder since it does not fulfill the Harris criterion~\cite{vojta2006rare}. Secondly, the imposed strain is a global constraint. That is, the sum of the local strains is constrained to the imposed value; this allows for banding. For contrast, the random organization model assumes that each particle experiences the same strain. Lastly, elasticity has effective quasi-long-range interactions which could affect the behavior~\cite{mari2022absorbing}. 

We have proposed a simple model that explains the occurrence of banding and the diverging distance between bands. The main assumption is that the energy at the largest strain is minimized. This is supported by simulations, which show that the energy decreases over time, especially near the transition. The minimization of energy could be an emergent organizing principle, similar to those proposed for driven materials in the dynamical regime~\cite{kedia2023drive,chvykov2021low}. Energy arguments have been previously used to explain shear bands in the quasistatic limit~\cite{dasgupta2012microscopic}.

The model predicts that the onset of plasticity occurs only above a threshold strain. The distance between bands diverges at the transition. The predicted exponents are different than those measured. This often occurs for mean-field approximations. The model does not capture the occurrence of plasticity below the transition. We note that the assumptions are minimal and not specific to bistable networks. The occurrence of banding, where the distance between bands diverges as one approaches the transition, may be a common feature for the oscillatory yielding transition. 

Lastly, we considered the polarization of the bonds that undergo a transition -- a new form of ordering. The elastic response is composed of an affine deformation and a non-affine contribution. Alignment with the imposed deformation is expected due to the affine contribution. However, with increased driving, the alignment increases. The time scales of the alignment are comparable to the relaxation time, suggesting a possible structural source of the long time scales. The alignment of the transitioning bonds provides a mechanism for encoding memory of the applied deformation. Furthermore, it is a mechanism for decreasing the energy cost of the deformation. This could be a crucial ingredient in understanding the occurrence of limit cycles.

\section*{Acknowledgments}
We would like to acknowledge Marc Berneman, Himangsu Bhaumik, Yoav Lahini, and Sefi Givli for insightful conversations. This work was supported by the Israel Science Foundation (grant 2385/20) and the Alon Fellowship.
    
\appendix

\section{Effect of the discretization of the strain steps}
\label{Appendix: steps per cycle}

\begin{figure}[h]
    \begin{center}
    \includegraphics[width=1\columnwidth]{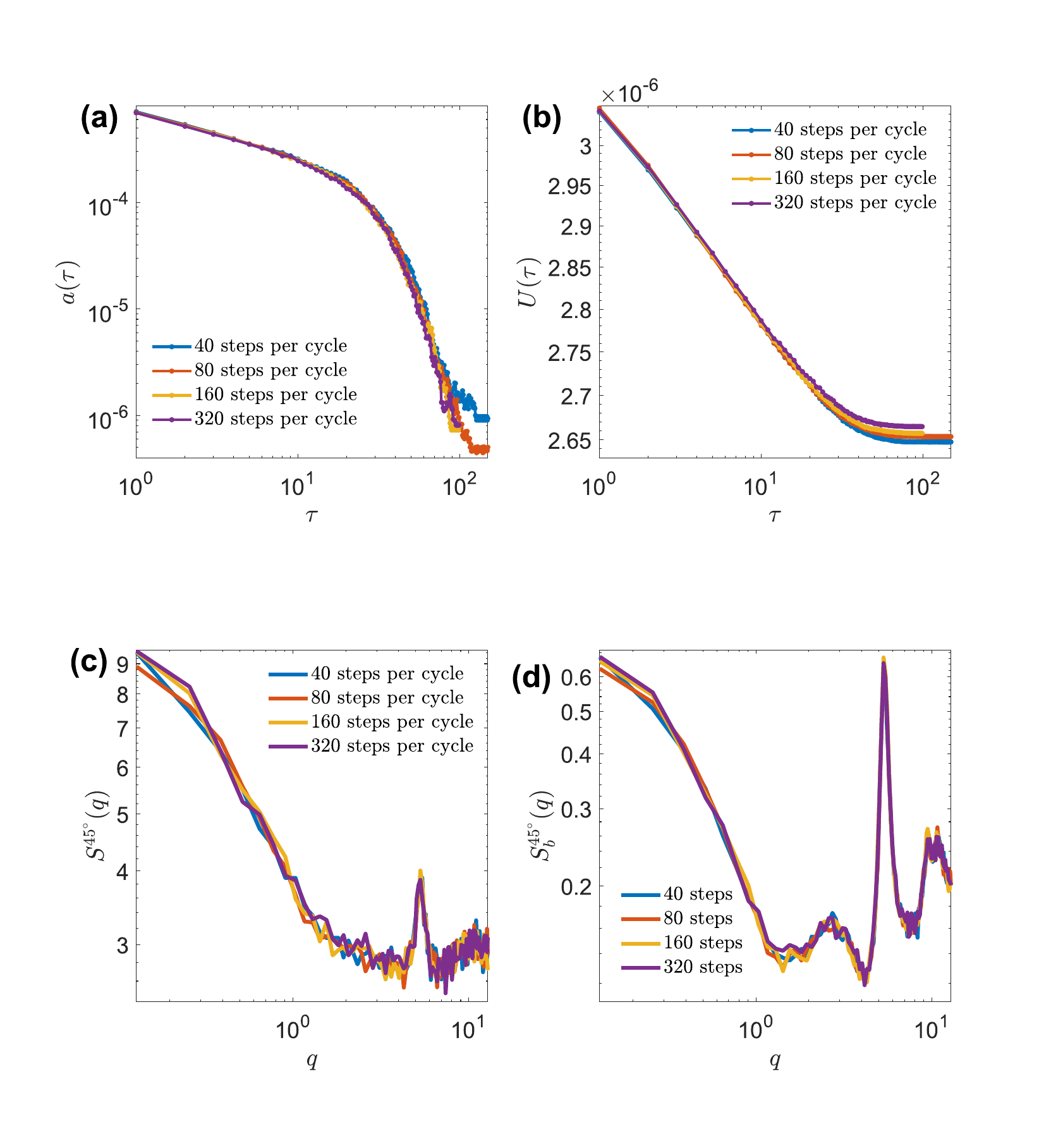}
    \end{center}
    \caption{\textbf{Effect of the strain size in a cycle} The size of the strain increment is given by, $\Delta \gamma =4\frac{\gamma_{max}}{steps}$. The discretization has little effect on our results. (\textbf{a}) Activity as a function of time. (\textbf{b}) Potential energy as a function of time. (\textbf{c}) Structure factor along $45^\circ$. (\textbf{d}) Structure factor of bond elongations in half a cycle along $45^\circ$. All data were obtained at amplitude $\gamma_{\max} = 0.012$, $N = 3000$ and $800$ realizations.}
    \label{fig:SQ-TT}
\end{figure}

To simulate quasistatic deformation, ideally, the strain step size should be smaller than the distance between instabilities. To simulate a large number of cycles, we are unable to have extremely fine discretization. This section of the appendix aims to investigate the effect of varying the step size of the strain. 
    
Figure~\hyperref[fig:SQ-TT]{{\ref{fig:SQ-TT}}} shows that varying the number of discretization steps per cycle has a very weak effect on the results.  Here, we examine four different step counts per cycle. Figures~\hyperref[fig:SQ-TT]{{\ref{fig:SQ-TT}}(a)} and \hyperref[fig:SQ-TT]{(b)} show the evolution of activity and potential energy, respectively. The curves are nearly identical, indicating that the system behaviors are insensitive to the number of steps in each cycle. In Figure~\hyperref[fig:SQ-TT]{{\ref{fig:SQ-TT}}(c)}, the structure factor measured along the \(45^\circ\) direction, corresponding to the orientation of instability bands, also shows no significant dependence on the number of steps. Similarly, the structure factor of bond elongations shown in Figure~\hyperref[fig:SQ-TT]{{\ref{fig:SQ-TT}}(d)} exhibits the same trend.

These results demonstrate that the effect of finite discretization in the shearing protocol is small and can be considered negligible when the number of steps per cycle exceeds 40. In our simulations, the maximum strain amplitude does not exceed 0.02, and the minimum number of steps per cycle is 40 (10 steps per quarter cycle). This results in a strain increment of less than \( 0.002 \) per step.

\section{Finite size effects near the transition}
\label{Appendix: finite size effects near the transition}

\begin{figure}[h]
    \begin{center}
    \includegraphics[width=1\columnwidth]{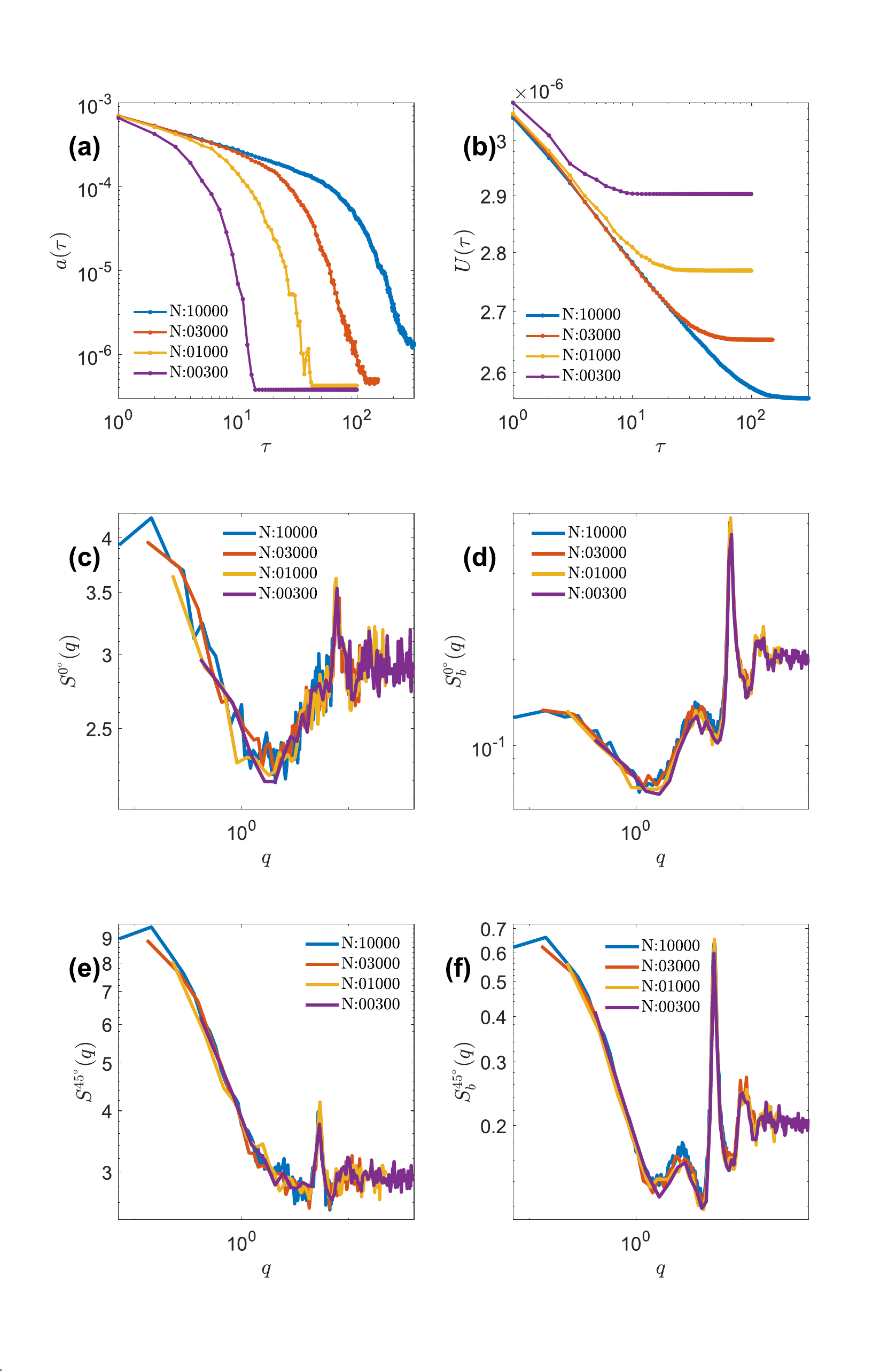}
    \end{center}
    \caption{\textbf{Finite size effects below the transition}. (\textbf{a}) shows the activity as a function of the number of cycles. (\textbf{b}) The evolution of the energy with the number of cycles. The energy decreases approximately logarithmically up to cut-off time, which grows with system size, and then approaches a plateau. This suggests logarithmic aging in the thermodynamic limit, $N\rightarrow{\infty}$. (\textbf{c})-(\textbf{f}) The structure factor for the transitioning bonds (left) and bond elongation (right) along $\theta=0^\circ$ (top) and $\theta=45^\circ$ (bottom). Overall, we do not observe significant finite-size effects. All data were obtained at amplitude $\gamma_{\max} = 0.012$.}
    \label{fig:SQ-NN}
\end{figure}

In this section, we consider the finite-size effects shown in Figure~\hyperref[fig:SQ-NN]{\ref{fig:SQ-NN}}, near but below the transition. Panel \hyperref[fig:SQ-NN]{(a)} shows the activity as a function of the number of training cycles. As discussed in the main text, the relaxation time grows as $N^{\frac{z}{2}}$ at the transition. In panel \hyperref[fig:SQ-NN]{(b)}, we show the evolution of the energy. The energy decreases approximately logarithmically, with a cutoff that grows with system size. Experiments on thin crumpled sheets have also observed aging effects~\cite{shohat2022memory,shohat2023logarithmic}. Activity allows the system to explore the state space and find lower energy states. The system ceases to evolve when the activity dies out.

Figure~\hyperref[fig:SQ-NN]{\ref{fig:SQ-NN}(c)} and \hyperref[fig:SQ-NN]{(e)} show the structure factor along $\theta=0^\circ$  and $\theta=45^\circ$ for the transitioning bonds. Figure~\hyperref[fig:SQ-NN]{\ref{fig:SQ-NN}(d)} and \hyperref[fig:SQ-NN]{(f)} shows the structure factor along $\theta=0^\circ$  and $\theta=45^\circ$ for the bond extension. In both cases, we do not observe significant finite-size effects.

\section{Supplementary Video}
\label{Appendix: Supplementary_Video:evolution_of_bands}

The supplementary video shows the evolution of bands over periodic drive when the system is above the transition. It can be accessed at {\color{blue} insert URL here}. The left panel shows the transitioning bonds, while the right-hand side shows the bond extension, comparing $-\gamma_{max}$ and  $\gamma_{max}$ (blue denotes bond elongation and red denoted compression). The cycle number is indicated in the top-left corner.%\href{https://drive.google.com/file/d/1H8MvgHuoM4T901JBKfqEoAed4-0Ccfnu/view?usp=sharing}{\texttt{here}}.
Note that there are distinct regions, which are essentially frozen and banded regions that evolve. There is also a population of transitioning bonds that hardly change from cycle to cycle.

\section{Snapshot of polarization of transitioning bonds}
\label{Appendix: V PO}

\begin{figure}[h]
    \begin{center}
    \includegraphics[width=.7\columnwidth]{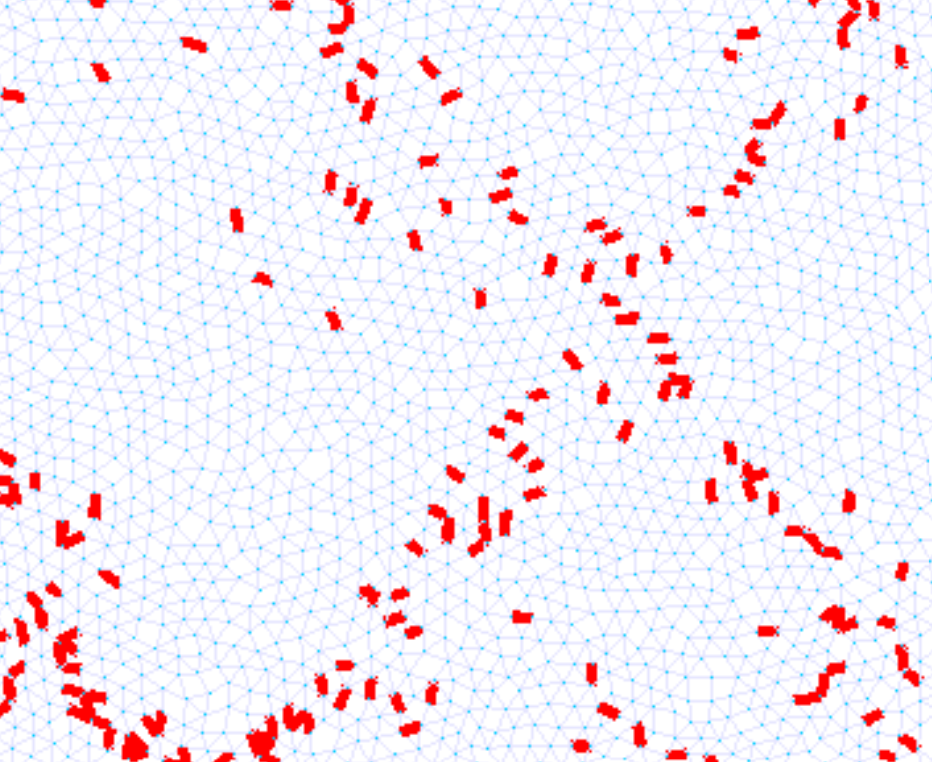}
    \end{center}
    \caption{\textbf{Polarization of transitioning bonds.} This panel shows a zoomed-in snapshot of Figure~\hyperref[fig:visualization-above-transition]{{\ref{fig:visualization-above-transition}}(b)}. The transitioning bonds (red) tend to align along $0^\circ$ and $90^\circ$, while the orientation distribution of all bonds (including both red and blue) is nearly uniform.}
    \label{fig:V PO}
\end{figure}

Here we show a zoom-in on the structure of the transitioning bonds in  Figure~\hyperref[fig:visualization-above-transition]{{\ref{fig:visualization-above-transition}}(b)}. The transitioning bonds shown in red are mostly aligned along the angles $\theta=0^\circ,90^\circ$.

\newpage

\newpage

\bibliographystyle{unsrt}
\bibliography{bibl}

\end{document}